\documentclass[a4paper,11pt]{article}
\usepackage{pos}

\title{Axial Anomaly, entanglement and polarization}

\author[a,b]{O,V. Teryaev}

\affiliation[a]{Joint Institute for Nuclear Research,\\
141980, Dubna, Russia}

\affiliation[b]{Moscow Center for Advanced Studies,\\
Kulakova str. 20, Moscow 123592, Russia
}

\emailAdd{teryaev@jinr.ru}
\emailAdd{oteryaev@gmail.com}

\abstract{ The (pion) decays  controlled by axial anomaly imply the
specific entanglement between photons having also  the counterparts for classical electromagnetic waves. This is also a specific case of Eisnstein-Podolsky-Rosen-Bohm-Aharonov effect. The absence of causality and non-locality in (angular) momentum conservation is manifested, being especially clear for
the generalization to the case of time rather than space separation
The similar decays in  external magnetic   field  manifest the interplay with vacuum conductivity in external magnetic field and longitudinal polarization of vector mesons observed in heavy-ion collisions.
Yet another interplay of polarization and fundamental symmetries is represented by the smallness of effective QCD coupling extracted from Bjorken sum rule and extended equivalence principle for spin-gravity interactions. }

\FullConference{QCD at the Extremes (QCDEX2025)\\
1–5 Sept 2025\\
Online\\}

\begin{document}
\maketitle
\section{Introduction}

The entanglement phenomena \cite{Einstein:1935rr,Bohr:1935af}
manifested in spin correlations of singlet states decay products \cite{Bohm:1957zz} remain of great interest for decades. Here I am pointing out to their relation to such phenomenon as axial anomaly
\cite{Adler:1969gk,Bell:1969ts}, whose co-discoverer, J.S. Bell, provided a major contribution \cite{Bell:1964kc} to the investigations of entanglement problem as well. This relation opens the complementary possibilities to study the fundamental problems of quantum theory in the
domain of high energy physics, in particular, in heavy-ion collisions.

I consider the  entanglement of decay photons polarizations in Section 2, comparing it also with the entanglement of time-separated photons in the crossing-related case of space-like virtual pions. Both cases are compatible with the picture of  the correlation emergence (corresponding, in particular, to the splitting processes in the framework of many-world interpretation of quantum  theory   \cite{Everett:1957hd}) implying the absence of causality. I also stress that the (angular) momentum conservation is essentially non-local.

The Section 3 is dedicated to the consideration of classical electromagnetic fields. If the decay photons are considered as
a plane classical waves, the entanglement corresponds to the correlation of electrical and magnetic fields in {\it different} waves, pronounced even if the waves are very far from one another. Another, most important, example, considered in Section 4, is the pion decay to lepton pair in the strong magnetic field (which, in particular, is generated in heavy-ion collisions). The entanglement analog in that case is the longitudinal polarization of the pair w.r.t. to magnetic field axis, which may be observed experimentally.
In the dense media, the role of magnetic field may be also played by vorticity coupled to anomaly via the chemical potential.
Moreover, the (tensor) polarization of vector mesons may be independent of magnetic field (and vorticity) and emerge due to spin correlations in their production process. It is pointed out that this effect is nothing else than entanglement  in time-reversed EPR (AB) process.

Section 5 contains discussion and  conclusions.

\section{Axial Anomaly and Entanglement}
\subsection{Pion decay}
Let us start with the expression for the (anomalous) VVP correlator of pseudoscalar current  with two vector ones, carrying momenta $q,k,p$ respectively:
\begin{equation}
\label{ampind}
M^{\mu \nu} = C(q^2)\varepsilon^{\mu \nu \alpha \beta}k_\alpha p_\beta \equiv C(q^2)\varepsilon^{\mu \nu k  p}.
\end{equation}
In the "rest" frame where q has only the time component the correlator has only the transverse components
\begin{equation}
\label{ampind}
M^{i j} = C(q^2)q^2\varepsilon^{i j}.
\end{equation}
It is related to decay amplitude of pseudoscalar state to two-photon with polarization vextors $e_1, e_2$  as
  \begin{equation}
  \label{ampol}
M^{P \to \gamma \gamma} = C \varepsilon^{e_1^* e_2^* k  p} = C q^2 \varepsilon^{e_1^* e_2^*}
\end{equation}
where coefficient $C$ now includes also the respective decay constant.
The square of this amplitude is
\begin{equation}
\label{dm}
|M^{P \to \gamma \gamma}|^2 = N (1-|(e_1^* e_2)|^2)  = N(1-Tr (\rho_1 \rho_2)),
\end{equation}
where $\rho_{ij}= e_i e_j^*$ is the density matrix of respective photon.
One can see that the measurement of linear polarization of one photon implies the definite value the second one in the orthogonal plane, while the
measurement of the definite circular polarization implies the opposite one of another photon, manifesting the EPR-AB effect. This property can be made especially
clear by rewriting expression in (\ref{dm}) as:
\begin{equation}
\label{dmpr}
(1-Tr (\rho_1 \rho_2))  =Tr ((1-\rho_1) \rho_2)= Tr ((1-\rho_2) \rho_1).
\end{equation}
 Note that the density matrices $\rho_{1,2}$ selected by the detector   is related (see e.g. \S 65 in \cite{Berestetskii:1982qgu}) to  the ones emerging in the scattering process ($\rho^f_{1,2}$) by the projection procedure defining the observable probability $dw$, which was also recently shown \cite{Teryaev:2022mpe}
to provide a sort of irreversibility:
\begin{eqnarray}
 \label{meas}
d w \sim Tr (\rho_a  \rho^{f}_a).
 \end{eqnarray}
Comparing this with (\ref{dmpr}) one get
\begin{equation}
\label{dmf}
 \rho_1^f= I-\rho_2, \,\   \rho_2^f= I-\rho_1.
\end{equation}
The measurement of the spin state of one photon immediately defines the state of another one {\it before} its measurement.
This is just the teleportation property which takes especially clear form in the case of scalar rather than pseudoscalar particle which will be discussed below.

Expressing the density matrices ({\ref{dmf}) in terms of Stokes parameters
\begin{equation}
\label{st}
 \rho_a =\frac{1}{2}(I + \sigma_i \xi_{i,a} ),
\end{equation}
one get
\begin{equation}
\label{stf}
\xi_{i,1}^f = - \xi_{i,2}, \,\  \xi_{i,2}^f = - \xi_{i,1}.
\end{equation}
The special role is played by circular polarization $\xi_2$ when (\ref{stf}) provides the (longitudinal) angular momentum conservation. The conservation is clearly non-local: when one of the photons reveals its angular momentum  after the measurement, the second one can be arbitrary far. Recall, that locality of conservation laws is deeply related \cite{feynchar} with dependence  of time ordering of the events (say, disappearance and appearance of charge)  on the choice of the reference frame. The similar situation with quantum measurement is  in fact the key issue in EPR  effect, where measurement with one of the particles will lead to the simultaneous (or even preceding in other
frames) appearance of definite state of another one,
so that the non-locality of  conservation laws is not too unnatural.
 
{\bf The natural implication of this situation is the {\it absence} of causal connection between the two measurements.
As there is no way to use the measurement for transfer of information, there is also no fundamental contradiction 
with causality. There is also no need to explain the entanglement through the correlation emerged in the initial state, so that the violation of the related Bell inequalities is quite natural.} 

Let us also stress that the angular momentum conservation is especially clear in the form (\ref{stf}). The measurement  of circular polarization
of one photon defines the final circular polarization of another one. At the same time, the measurement of linear polarization of the last photon
provides the probability $1/2$, according to (\ref{dm}).
Such a measurement of linear polarization, when circular one is defined by angular momentum conservation may be considered as a manifestation of the general problem formulated in the coinciding titles of \cite{Einstein:1935rr,Bohr:1935af} (Can quantum mechanical description of physical reality be considered complete?). As the linear polarization appears in the projection of circular one (so that they are not measured simultaneously), the positive answer of \cite{Bohr:1935af} to that question is actually implied here \footnote{To be more close to the problem posed in \cite{Einstein:1935rr} one may consider the different components of
Dirac fermions polarization instead of linear and circular polarization of photons, so that the impossibility of their simultaneous measurement is
compatible with the non-commutativity of corresponding operators}.

One may also  extend the calculation of polarization produced in the scattering process to the second photon. The simplest way is to square the  amplitude {\ref{ampind}) with free indices, as the polarization vectors describe the measurement process. The resulting normalised density matrix is
\begin{equation}
\label{sc2}
 \rho^{a,b}_{ij,kl} =\frac{1}{2}(\delta_{ij} \delta_{kl}-
\delta_{il} \delta_{kj}).
\end{equation}
This expression describes in the clear way the entanglement of photons as identical \cite{Benatti:2020xgb} particles.
The presence of identical photons results in the antisymmetry
(compensated  by antisymmetry w.r.t. momenta interchange)
in interchanges $$i \leftrightarrow k$$
and
$$j \leftrightarrow l$$.

\subsection{Scattering on the $t$-channel pion}
The non-locality of conservation laws and absence of naive causality is even stronger pronounced in the crossing related process of the photon scattering on the (space-like) pion. The latter may correspond to the Compton scattering on the proton mediated by pion pole,
which in the case of observation  of pion in the final state is known to be the way to measure of pion formfactor (Sullivan process).
The scattering amplitude takes the form
\begin{equation}
\label{ampolt}
M^{\gamma P \to  \gamma} = C \varepsilon^{e_1 e_2^* k  p} = C q^2 \varepsilon^{e_1 e_2^*},
\end{equation}
leading after squaring to
\begin{equation}
\label{dms}
|M^{ \gamma P \to  \gamma}|^2 = N (1-|(e_1 e_2)|^2)  = N(1-Tr (\rho_1^T \rho_2)).
\end{equation}
This corresponds to the expression for density matrix of the final photon in terms of the (prepared) initial one
\begin{equation}
\label{dmfs}
 \rho_2^f= I-\rho_1,
\end{equation}
corresponding to its Stokes parameters
\begin{equation}
\label{stfs}
\xi_{i,2}^f = - \xi_{i,2} (i=1,3), \,\  \xi_{2,2}^f = \xi_{2,1}.
\end{equation}
These relations express the properties of polarization plane rotation (for $\xi_{1}, \xi_{3}$) and angular momentum conservation (for $\xi_{2}$).

The EPR-AB effect can now have a new properties of retrodiction due to crossing invariance.


Let us consider the preparation of the state $1$ in the same process of pion decay considered earlier, so that
combination of $s$ and $t-$ channel processes looks like
\begin{equation}
\label{st}
\pi(s) \to \gamma_1+\gamma_3; \gamma_1+\pi(t) \to \gamma_2.
\end{equation}
As now both (\ref{stf}, \ref{stfs}) are valid, and measurement  of circular polarization of the photon $2$ leads to definite polarization of both $3$ (separated in space) and $1$ (separated in time and, moreover, existed in {\it past}). Let us stress that the conservation of angular momentum in each event is crucial.

If the scattering on the t-channel pion is considered in the Breit frame (where photon reverse its momentum conserving the energy) the measurement of the photon $3$ polarization may be performed in the same laboratory, where first pion decayed some time ago, affecting the polarization produced in that process in past. This measurement fixes also the polarization of photon $3$ separated by space-like interval, like in usual EPR-AB process.

Such possibility of affecting the past seems to be generic for EPR-AB processes. To be more close to the original AB version, one may consider the decay of scalar particle (say, Higgs boson) to the pair of unstable spin $1/2$  particles (say, $\Lambda \bar \Lambda$), decaying, in turn, to spin $1/2$ and spin $0$ particles ((anti)proton and pion), so that the analog of (\ref{st}) looks like:
\begin{equation}
\label{stAB}
H \to \Lambda + \bar \Lambda; \Lambda \to p + \pi^- ; \bar \Lambda \to \bar p + \pi^+.
\end{equation}
The measurement of spin of proton leads not only to the selection of that of (distant) antiproton,
but also of already decayed $\Lambda$ and $\bar \Lambda$.

The decay of scalar particles also manifests in the most clear way the "teleportation" of the measured polarization (say, of antiproton) to  the emerged polarization of other decay product (in that case, proton) before measurement.
To do so it is sufficient to compare the expression for decay width to proton and antiproton with momenta $p_1,p_2$ and measured polarizations $P_1^d , P_2^d$, respectively,
\begin{equation}
\label{Htele}
d \Gamma \sim Tr [\rho^p (p_1,P_1^d) \rho^{\bar p} (p_2,P_2^d)].
\end{equation}
to the relation between the measured $P_1^d$ and emergent  $P_1^f$ polarizations of proton,
\begin{equation}
\label{Hteled}
d \Gamma \sim Tr [\rho^p (p_1,P_1^d) \rho^p (p_1,P_1^f)].
\end{equation}
One can see that the measurement corresponds to the  teleportation to antiparticle. To perform the quantitative comparison one should consider  (\ref{Htele}) in the center of mass frame
\begin{equation}
\label{Htelecm}
d \Gamma \sim 1- \vec P_1^d \vec P_2^d,
\end{equation}
while  (\ref{Hteled}) in the rest frame of proton is
\begin{equation}
\label{Hteledr}
d \Gamma \sim 1- \vec P_1^d \vec P_1^f,
\end{equation}

To avoid the contribution of orbital angular momentum, one  may consider the case of all spins and momenta directed along the same axis. This corresponds to particles of definite helicity. The measurement of, say,  helicity $+1$ for the proton results, due to angular momentum conservation, in the same helicity $+1$ for antiproton, $\Lambda$ and $\bar \Lambda$.

Note also that in this case the possible specifics due
to the interplay of entanglement and indistinguishability \cite{Benatti:2020xgb} is absent.

In general, the consideration of  the crossed version of EPR-AB experiment makes the violation of causality in
measurement process (and in the Universe splitting in the manyworld interpretation) even more clear than in usual case, where it is manifested by applying the Lorentz invariance.

\section{Classical fields and entanglement}

The entanglement is essentially quantum notion. At the same time, the photons may be considered also as a classical plane waves. As the electric and magnetic fields are than mutually orthogonal, one may wonder, how the only possible
pseudoscalar invariant $\vec E \vec H$ can be formed.

This seeming contradiction is resolved by inspection of the matrix element (\ref{ampol}). As the polarization vectors correspond to the Fourier transforms of the respective potentials, it corresponds to the scalar products of field strength in {\it different} waves
\begin{equation}
M^{P \to \gamma \gamma} = C \varepsilon^{e_1^* e_2^* k  p} \sim \vec E_1 \vec H_2 + \vec E_2 \vec H_1.
\end{equation}
This expression correspond to the correlation between the polarization planes of two arbitrary distant waves, keeping the $90^0$ angle
between them. This is a classical counterpart of entanglement.

The similar result holds also for $t-$channel scattering amplitude (\ref{ampolt}) where it corresponds to well-known classical phenomenon of the rotation of polarization plane, being the counterpart of the time-separated  entanglement.

\section{Axial anomaly, magnetic field and tensor polarization of time-like  photons}

The recent discovery of strong tensor polarization of vector mesons in heavy-ion collisions
\cite{Acharya:2019vpe} attracted major attention to the mechanisms of its generation and its relation to the extremely large vorticity of strongly interacting matter \cite{Chen:2020pty, Weickgenannt:2020aaf, Fukushima:2020yzx, Kapusta:2020dco,          Sheng:2020ghv, Sheng:2019kmk}.

The tensor polarization is P-even quantity which is proportional to the product
\cite{Efremov:1981vs, Xu:2003fq} of uncorrelated (anti)quark polarizations constituting the vector meson.

The (P-odd) quark polarization transferred to hyperon polarization is rather small. In the approach\cite{Rogachevsky:2010ys} based on hydrodynamic approximation for axial anomaly \cite{Son:2009tf,Sadofyev:2010is} the rapid growth of polarization for smaller energies was predicted. Assuming the quark-hadron duality in terms of axial charge and applying the kinetic Quark-Gluon String Model the $\Lambda$ polarization in $Au-Au$ collisions and $\sqrt{s} = 5 GeV$ was estimated   \cite{Baznat:2013zx} to be at the order of $1\%$. This estimate later appeared to be in semi-quantitative agreeement with seminal STAR data \cite{STAR:2017ckg}. The smallness of polarization and its decrease with energies was also confirmed  \cite{Becattini:2016gvu} in  the detailed calculations in the thermodynamic \cite{Becattini:2009wh} framework. The numerical similarity \cite{Baznat:2015eca,Ivanov:2020qqe} of the results obtained in anomalous  (Axial Vortical Effect, AVE) and thermodynamic approaches may be not a coincidence but rather yet another manifestation of quark-hadron duality  \cite{Prokhorov:2017atp, Sorin:2016smp,Teryaev:2017wlm,Prokhorov:2019yka}.

As the data on tensor polarization \cite{Acharya:2019vpe} are exceeding the naive
estimate of vector polarization squared, this requres some explanation. Recently, this discrepancy was attributed \cite{Sheng:2019kmk} to the collective $\phi-$meson field.
At the same time, the relation between (anti)quarks polarizations and vector meson tensor polarization, as originally derived in  \cite{Efremov:1981vs} includes also {\it correlation} of polarizations. In perturbation theory it emerges already at Born level and may be non-zero if (anti)quark polarizations are arbitrary small. In particular, the $\rho$ tensor polarization in $p \bar p$ collision was explained \cite{Efremov:1982mr} in the fusion model due to the vector coupling between quark and antiquark.
Note that such correlation may be also considered as a sort of entanglement, now reversed in time,

The role of vector current was also manifested in the  relation \cite{Bratkovskaya:1995kh} of tensor polarization to the tensor structures in the correlator of quark currents which was introduced \cite{McLerran:1984ay} for the calculation of dilepton (and photon) production rates.

The similar obiect appears in the description of vacuum conductivity and its component parallel to magnetic revealed in lattice calulations \cite{Buividovich:2010tn} was related \cite{Buividovich:2012kq}   to {\it longitudinal} (wrt magnetic field direction) polarization of soft dileptons. The lattice calculations were recently performed \cite{Astrakhantsev:2019zkr} in more realistic framework and with higher accuracy and the dominance of longitudinal conductivity
was confirmed and found to be growing with magnetic field manifesting dynamical Chiral Magnetic Effect (DCME).    The similar effect was also found for Dirac semimetals \cite{Ulybyshev:2013swa}.

 The vector correlators in the external magnetic field were later studied on the lattice in connection with  vector (mostly $\rho$) mesons properties \cite{Luschevskaya:2014lga,Luschevskaya:2016epp} and the dominance of longitudinal polarization was observed also in that case.

To clarify the anomaly contribution to  vector meson polarization one should consider the  amplidude of pseudoscalar meson decay in external electromagnetic field
\begin{equation}
\label{ampolt}
M^{{\bf  H} (k) P \to  V(p)} = C(M_P) \varepsilon^{A(k) e_V^* k  p},
\end{equation}
where $e_V$ is a polarization vector of virtual photon
 In the meson rest frame only the magnetic field contribute
\begin{equation}
\label{ampolt}
M^{{\bf  H} (k) P \to  V(p)} = C(M_P) M_P (\vec{e^*}_V  \vec H),
\end{equation}
and the longitudinal polarization of virtual photon with respect to magnetic field direction is clear.

\section{Discussion and Conclusions}
The axial anomaly naturally implies the covariant description of entanglement.  It allows one to observe the absence of causality
in the quantum measurement procedure, including the action on the past,  which does not lead  to information  transfer and any of related paradoxes.

Anomaly in external magnetic field explains the longitudinal polarization of virtual photons in heavy-ion collisions. Interestingly, the measurement of the polarization of lepton produced in the direction orthogonal to magnetic field allows one to determine the projection of this photon polarization to this direction, realizing the mentioned action on the past.

The role of fundamental symmetries in the polarization effect is also manifested in the smallness of effective QCD constant 
	extracted from the polarized Bjorken sum rule \cite{Deur:2023dzc}.  In fact, this smallness is coming because the Gerasimov-Drell-Hearn sum rules for proton and neutron are very close. This, in turn, is due to the closeness of the squared anomalous magnetic moment of proton and neutron, which may be related to the extension 
\cite{Teryaev:1999su,Teryaev:2016edw} of Equivalence Principle for spin-gravity interactions, resulting in its validity for quarks and gluons separately. As a result, extra-dimensional gravity, rooted in approximate conformality, appears to be related to the real one.  

This work was supported by the Russian Science Foundation (Grant No. 25-72-30005).

\end{document}